\documentclass[12pt,english,american,preprint]{article}
\usepackage[latin9]{inputenc}
\usepackage{color}
\usepackage{array}
\usepackage{float}
\usepackage{multirow}
\usepackage{amsmath}
\usepackage{amssymb}
\usepackage{graphicx}

\newcommand{\eqb}{\begin{equation}}
\newcommand{\eqe}{\end{equation}}
\newcommand{\dmb}{\begin{displaymath}}
\newcommand{\dme}{\end{displaymath}}

\newcommand{\eab}{\begin{eqnarray}}
\newcommand{\eae}{\end{eqnarray}}

\newcommand{\be}{\begin{equation}}
\newcommand{\ee}{\end{equation}}

\begin{document}
\begin{titlepage}

\begin{center}
\Large{Relic photon temperature versus redshift and \\ 
the cosmic neutrino background}\vspace{1.5cm}\\ 
\large{Ralf Hofmann}
\end{center}
\vspace{2.0cm} 
\begin{center}
{\em Institut f\"ur Theoretische Physik\\ 
Universit\"at Heidelberg\\ 
Philosophenweg 16\\ 
69120 Heidelberg, Germany}
\end{center}
\vspace{1.0cm}
\begin{abstract}

\end{abstract} 
Presuming that CMB photons are described by the deconfining phase of an SU(2) Yang-Mills theory with the 
critical temperature for the deconfining-preconfining phase transition matching the present 
CMB temperature $T_0\sim 2.725\,$K (SU(2)$_{\tiny\mbox{CMB}}$), we investigate how CMB temperature $T$ connects with the 
cosmological scale factor $a$ in a Friedmann-Lema\^itre-Robertson-Walker Universe. Owing to a violation of conformal scaling at late times, 
the tension between the (instantaneous) redshift of reionisation from CMB observation ($z_{\tiny\mbox{re}}\sim 11$) and 
quasar spectra ($z_{\tiny\mbox{re}}\sim 6$) is repealed. Also, we find that the redshift of CMB decoupling moves from $z_{\tiny\mbox{dec}}\sim 1100$ 
to $z_{\tiny\mbox{dec}}\sim 1775$ which questions $\Lambda$CDM cosmology at high redshifts. 
Adapting this model to the conventional physics 
of three flavours of massless cosmic neutrinos, we demonstrate inconsistency with the value 
N$_{\tiny\mbox{eff}}\sim 3.36$ extracted from Planck data. Interactions between cosmic neutrinos and the 
CMB implies a {\sl common} temperature $T$ of (no longer separately conserved) 
CMB and neutrino fluids. N$_{\tiny\mbox{eff}}\sim 3.36$ then entails a universal, temperature induced cosmic 
neutrino mass $m_\nu=\xi T$ with $\xi=3.973$. Our above results on $z_{\tiny\mbox{re}}$ and $z_{\tiny\mbox{dec}}$, 
derived from SU(2)$_{\tiny\mbox{CMB}}$ alone, are 
essentially unaffected when including such a neutrino sector.

\end{titlepage} 

\section{Introduction} 

Cosmology has become a science of precision data in all its main experimental branches: 
large-scale structure surveys, e.g. \cite{SLDSSI,SLDSSII}, observations of the cosmic microwave 
background (CMB), e. g. \cite{Cobe,WMAP,PlanckO}, 
and use of calibrated standard candles for luminosity distance - redshift measurements, e. g. \cite{Spergel,Riess}. Thanks to this 
fortuitous observational situation, we appear to possess an accurate parametrisation of the Universe's composition, 
and we understand how the CMB decoupled, what the statistical properties of its temperature fluctuations are, 
how matter structure grew, and what its late-time effects on the propagation of the CMB are. 
The present $\Lambda$CDM concordance model \cite{PlanckCP}, stating that apart from $\sim\,$70\% dark energy the 
present Universe is composed of 30\% nonrelativistic matter of which about 5.5\% 
is baryonic matter, is an apparently good one. Yet, $\Lambda$CDM  merely represents a simple parametrisation of cosmological expansion with no definite, falsifiable handle on what 
the dark sector in it actually represents. Moreover, certain signatures or parameter values, such as 
the redshift of reionisation or the present value of the Hubble parameter, are 
at considerable tension \cite{PlanckCP}, and 
there are degeneracies when exclusively relying on one 
observational modality. This precludes the use of another modality as an 
independent check. Finally, the CMB behaves in an anomalous way at large angles \cite{PlanckAno}: CMB cold spot 
\cite{CS}, large-angle suppression of the temperature-temperature (TT) correlation function on the ecliptic north, 
alignment of low CMB multipoles \cite{LA}, etc. This startling state of affairs suggests that, in contrast to a highly developed and efficient 
computational and statistical machinery, our present {\sl theories} of the matter, radiation and dark energy 
content of the Universe, and therefore of the cosmological model, are incomplete. 

In this work we propose a potential improvement of this situation by first addressing the 
late-time and then the high-redshift consequences of the postulate that 
photon propagation is, fundamentally, described by a pure SU(2) Yang-Mills 
theory of scale $\sim 10^{-4}$\,eV, SU(2)$_{\tiny\mbox{CMB}}$, see \cite{HofmannBook2012} and 
references therein. In doing so, we will conclude that at low redshift the 
conventional, conformal scaling law $T=a^{-1}\,T_0$ of the CMB temperature $T$ ($T_0=2.725\,$K the 
present CMB temperature; $a$ the cosmological scale factor, defined to be unity at present) 
is violated. In particular, we obtain $T(a=1/10)=6.2\,T_0$ with conformal scaling $T\propto a^{-1}$ essentially being 
restored for $a<1/10$. Relying on N$_{\tiny\mbox{eff}}\sim 3.36$ \cite{PlanckCP}, 
this implies consequences for cosmic neutrinos, first addressed by assuming their masslessness and subsequently by 
invoking a particular mechanism for mass generation by their interaction with 
the CMB on cosmological time scales. 

This report is organised as follows. In the next section we 
review and explicate properties of deconfining SU(2)$_{\tiny\mbox{CMB}}$ on the free quasiparticle level which are 
relevant for cosmology: thermodynamical quantities, evolution of effective gauge 
coupling, equation of state, and the deconfining-preconfining transition temperature 
$T_c$ for SU(2)$_{\tiny\mbox{CMB}}$. We also remark on the thermal photon's polarisation tensor 
in SU(2)$_{\tiny\mbox{CMB}}$, representing the by far dominant one-loop correction to 
free propagation. Apart from their role in thermalising photons to the 
temperature of the thermal ground state and its massive quasiparticle 
excitations, radiative corrections are, however, irrelevant to the cosmological questions 
addressed in the present work. In Sec.\,\ref{tvsahighz} we explore implications of 
SU(2)$_{\tiny\mbox{CMB}}$ by studying conservation of its energy in an expanding 
Friedmann-Lema\^itre-Robertson-Walker Universe. We find that conformal scaling 
of $T$ vs. $a$ is violated at low but restored at high redshifts. This resolves 
the tension between the high and low values of the redshift $z_{\tiny\mbox{re}}$ associated with instantaneous reionisation and extracted 
from CMB data \cite{PlanckCP} and quasar spectra \cite{Quasar}, respectively. 
On the other hand, the value of redshift $z_{\tiny\mbox{dec}}$ for CMB decoupling is increased by a 
factor $\sim 1.6$ which challenges the validity of the $\Lambda$CDM concordance 
model at high redshifts. In Sec.\,\ref{Masslessneutr} we confront SU(2)$_{\tiny\mbox{CMB}}$ with conventional, massless 
cosmic neutrinos to conclude that $N_{\tiny\mbox{eff}}=3.36$ \cite{PlanckCP} is incompatible with three 
flavours. This situation changes if neutrinos are assumed to interact with the 
CMB, thus acquiring the same temperature and a temperature dependent mass. Importantly, such coupling of the 
neutrino sector to the CMB, motivated by the interpretation of a neutrino as a single center-vortex 
loop in the confining phase of an SU(2) Yang-Mills theory \cite{Moosmann2008}, does not in any essential way
change the conclusions of Sec.\,\ref{tvsahighz} on the values of $z_{\tiny\mbox{re}}$ and $z_{\tiny\mbox{dec}}$. 
Finally, we present a summary and our conclusions in Sec.\,\ref{Sum}. 
We also speculate about changes implied by SU(2) $_{\tiny\mbox{CMB}}$ and by the thus 
modified neutrino sector for the matter sector of the cosmological model at high redshift.          

This work employs natural units: Boltzmann's constant $k_B$, the reduced quantum of action $\hbar$, 
and the speed of light in vacuum $c$ are all set equal to unity.    

\section{SU(2)$_{\tiny\mbox{CMB}}$: Thermal ground state plus free quasiparticle excitations\label{SU(2)cmb}}

For the reader's convenience we review here results on deconfining SU(2) Yang-Mills thermodynamics relevant to 
the present work, for more detailed presentations see 
\cite{HofmannBook2012} and references therein.  

\subsection{Pressure and energy density\label{Pandrho}}  

In its deconfining phase an SU(2) Yang-Mills theory develops a thermal ground state invoked by 
spatially coarse-grained Harrington-Shepard calorons and anticalorons \cite{HS1977}, that is, temporally periodic instantons of 
trivial holonomy and topological charge of modulus unity. These field configurations are constructed in singular gauge thanks to a 
superposition principle for their prepotential discovered in \cite{'thooft1976}. (Anti)\-calorons 
are (anti)\-selfdual gauge field confi\-gurations. As such they exhibit {\sl vanishing} energy density and pressure. 
Their spatially coarse-grained reincarnation, an adjoint scalar field $\phi$, thus is inert. It describes part of the thermal ground state: 
there are no quantum fluctuations or classically propagating modes of this field. Fundamentally, propagating gauge field 
fluctuations, which are (slightly) harder than $|\phi|$, lift the energy density $\rho_{\tiny\mbox{gs}}$ of the thermal ground state from zero to
\eqb
\label{groundstate ed}
\rho_{\tiny\mbox{gs}}=4\pi\Lambda^3 T\,,
\eqe
where $\Lambda$ denotes the Yang-Mills scale (a free parameter of mass dimension one). After coarse-graining 
this phenomenon is associated with a simple pure-gauge configuration $a^\mu_{\tiny\mbox{gs}}$. For the ground-state 
pressure $P_{\tiny\mbox{gs}}$ one has $P_{\tiny\mbox{gs}}=-\rho_{\tiny\mbox{gs}}$. Field $\phi$ implies an 
adjoint Higgs mechanism. As a consequence, two of the three 
base directions of the SU(2) Lie algebra become massive while the third one, the direction 
identified with the photon, remains massless. Namely, in unitary gauge $\phi^a=\delta^{a3}\,|\phi|$ ($a=1,2,3$) 
a common mass $m=2e\sqrt{\frac{\Lambda^3}{2\pi T}}$ emerges for direction 1 and 2 which thus become thermal 
quasiparticles. Here $e$ is the effective gauge coupling whose temperature dependence is yet to be determined. 
One can show that the transition from the gauge, where $\phi$'s dynamics was derived (winding gauge), 
to unitary gauge transforms the Polyakov loop on $a^\mu_{\tiny\mbox{gs}}$ by an electric 
center jump, meaning that the thermal ground state is ${\bf Z}_2$ degenerate. 
This, in turn, implies the deconfining nature of the thermal ground state. 
On the level of free quasiparticles (one loop), one obtains the following expressions 
for the deconfining-phase pressure $P$ and energy density $\rho$ which both are sums of a ground-state and 
free thermal-fluctuation contributions (free quantum fluctuations are negligibly small): 
\eab
\label{Ponelooprhooneloop}
P(\lambda)&=&-\Lambda^4\left\{\frac{2\lambda^4}{(2\pi)^6}\left[2\bar{P}(0)+6\,\bar{P}(2a)\right]+2\lambda\right\}\,,\nonumber\\ 
\rho(\lambda)&=&\Lambda^4\left\{\frac{2\lambda^4}{(2\pi)^6}\left[2\bar{\rho}(0)+6\,\bar{\rho}(2a)\right]+2\lambda\right\}\,,
\eae
where 
\eab
\label{defdimPdimrho}
\bar{P}(y)&\equiv&\int_0^\infty
dx\,x^2\,\log\left[1-\exp(-\sqrt{x^2+y^2})\right]\,,\nonumber\\  
\bar{\rho}(y)&\equiv&\int_0^\infty
dx\,x^2\frac{\sqrt{x^2+y^2}}{\exp(\sqrt{x^2+y^2})-1}\,,
\eae
\eqb
\label{defamass}
a\equiv\frac{m}{2T}\,,
\eqe
and $\lambda\equiv\frac{2\pi T}{\Lambda}$. One may wonder why the contribution of the massless mode 
(referred to as $\gamma$ in the following) is thermalised to the 
same temperature $T$ as the tightly ground-state coupled massive, propagating fields (Higgs mechanism!), 
in the following referred to as $V^\pm$. (No coupling between $\gamma$ and $V^\pm$ takes place 
on the level of Eq.\,(\ref{Ponelooprhooneloop}).) To understand this, the results of 
\cite{Schwarz2007} are required where radiative corrections to the pressure are computed. While two-loop corrections generally rise with a lower power in $T$ than four, 
the $\gamma$-$V^\pm$ 2-loop diagram with a single four-vertex represents an exception. This two-loop diagram 
{\sl is} $\propto T^4$ and thus induces a weak coupling between $\gamma$ and $V^\pm$ 
for all temperatures in the deconfining phase. (The critical temperature $T_c$ for the deconfining-preconfining phase transition is not really an 
exception because there $\gamma$ starts to acquire an effective mass by tunneling between the deconfining and 
preconfining ground states \cite{radioHofmann2009}.) Notably on cosmological time scales one thus is 
assured that $\gamma$ and $V^\pm$ are thermalised to one and the same temperature. The (negative) 
contribution of this particular radiative correction was exploited in \cite{SpatialWilson2009} to infer the density 
of invisible magnetic monopoles and antimonopoles, liberated by rare (anti)caloron dissociation due 
to large holonomy shifts \cite{Nahm,VanBaal,LeeLu,Diankonov}. 
Effectively but weakly, this reduces the photon temperature. 
For our present purposes we may safely ignore this tiny effect (fraction of a per mille) \cite{SpatialWilson2009}.    

\subsection{Effective gauge coupling\label{effe}}

\begin{figure}
\begin{center}
\leavevmode
\leavevmode
\vspace{4.9cm}
\includegraphics{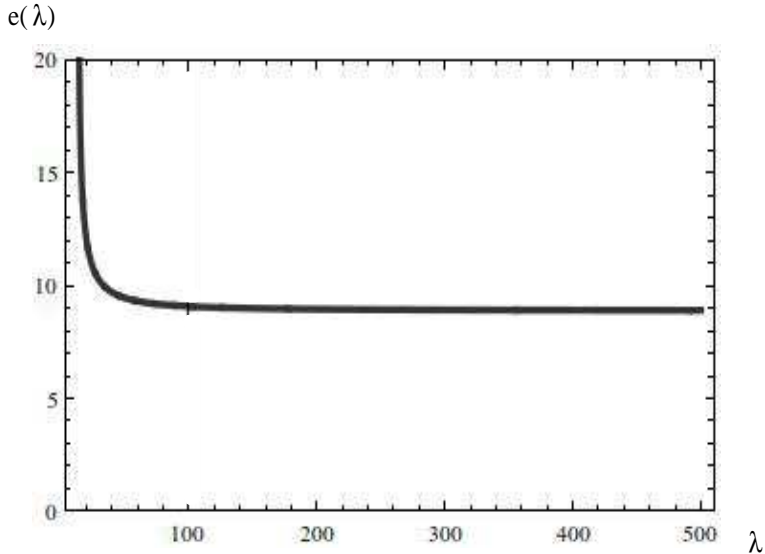}
\end{center}
\caption{\protect{\label{Fig-1}} Temperature dependence of effective coupling $e$. For $\lambda\searrow\lambda_c\equiv 13.78$ one 
has $e\propto -\log(\lambda-\lambda_c)$ while $e$ rapidly approaches the plateau $e=\sqrt{8}\pi$ as $\lambda$ 
increases away from $\lambda_c$.}      
\end{figure}

$P$ and $\rho$ in Eq.\,(\ref{Ponelooprhooneloop}) depend on the effective gauge coupling $e$ via the $V^\pm$ mass $m$, 
given as $m=2e\sqrt{\frac{\Lambda^3}{2\pi T}}$ or $\frac{m}{2T}\equiv a=2\pi e\lambda^{-3/2}$. To determine $e(\lambda)$, 
one thus requires $\lambda(a)$ or $a(\lambda)$. Thermodynamical consistency of the expressions in 
Eq.\,(\ref{Ponelooprhooneloop}), is equivalent to $\lambda(a)$ obeying the following first-order ordinary differential 
equation
\begin{equation}
\label{evoleqsu2}
\partial_a\lambda=-\frac{24\lambda^4
  a}{(2\pi)^6}\frac{D(2a)}{1+\frac{24\lambda^3a^2}{(2\pi)^6}D(2a)}\,,
\end{equation}
whose solution $\lambda(a)$ can be inverted to $a(\lambda)$ because 
\eqb
\label{D(y)}
D(y)\equiv \int_0^\infty dx\,\frac{x^2}{\sqrt{x^2 + y^2}}\,\frac{1}{\exp(\sqrt{x^2+y^2})-1}>0\,, \ \ \ \ (y\ge 0)\,.
\eqe
There is an attractor solution to Eq.\,(\ref{evoleqsu2}) which predicts a plateau at high temperatures,  
 $e=\sqrt{8}\pi$, and critical behaviour $e\propto -\log(\lambda-\lambda_c)$ just above 
$\lambda_c=13.87$. Fig.\,\ref{Fig-1} depicts function $e(\lambda)$.  
Because of the constancy of $e$ at high temperatures the $V^\pm$ mass $m$ drops like $\propto 1/\sqrt{T}$ as temperature rises. 
At sufficiently high temperatures we are thus facing a gas of $2+2\times3=8$ 
(3 polarisations for $V^\pm$) relativistic 
degrees of freedom. (The ground state contributions to energy density and pressure are only linearly rising 
in temperature. Thus they can be neglected at high temperatures.)  

Strictly speaking, given a temperature $T$, thermodynamics takes place on the 
one-loop level only: small radiative corrections to the free quasiparticle pressure are usually  
thermodynamically inconsistent in the sense that they do not obey the usual 
Legendre transformations. However, because a particular radiative correction to the pressure, mentioned in Sec.\,\ref{Pandrho}, is 
$\propto T^4$ the associated decrease in the pressure of 
free and massless thermal fluctuations (photons), introduced by a slight drop of temperature, can be interpreted as 
a decrease in photonic energy density related to that of the pressure by a Legendre transformation \cite{SpatialWilson2009}. 

\subsection{Equation of state\label{eossu2cmb}}

\begin{figure}
\begin{center}
\leavevmode
\leavevmode
\vspace{4.9cm}
\includegraphics{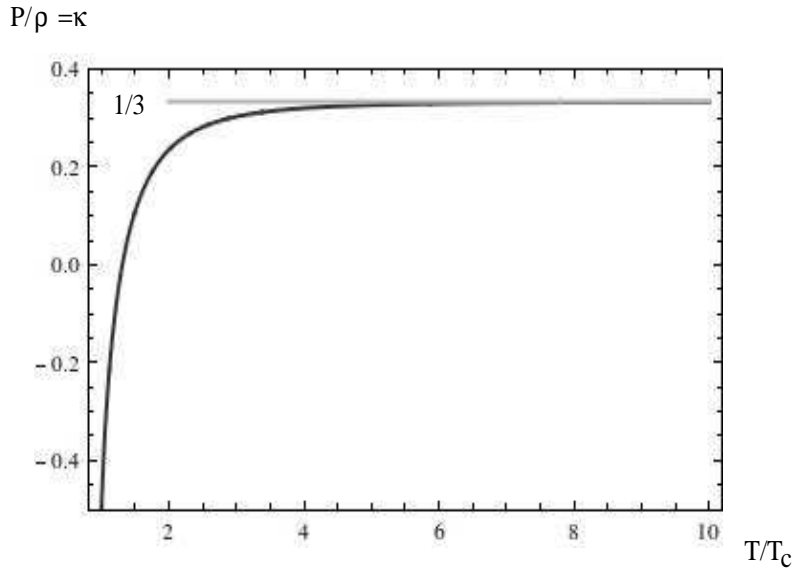}
\end{center}
\caption{\protect{\label{Fig-2}} Ratio of pressure and energy density, or equation-of-state parameter $\kappa$, 
as a function of $T/T_c$ in the deconfining phase of 
SU(2) Yang-Mills thermodynamics. Note the rapid approach to the behaviour of a thermal gas of 
free and massless particles ($P=\frac13\,\rho$). }      
\end{figure}

Let us now investigate how rapidly the parameter $\kappa$ in a presumed equation of state\footnote{By definition, 
the actual equation $P=P(\rho)$ of state, $P(\rho)$
being a nonlinear function, exhibits no 
explicit $T$ dependence, see also Sec.\,\ref{eneconsSEC}.} of the form  $P=\kappa\rho$ 
approaches that of a thermal gas of massless particles of 
deconfining SU(2) Yang-Mills thermodynamics as 
temperature increases away from $T_c$. Fig.\,\ref{Fig-2} depicts $\kappa$ as a function of $T/T_c$. 
Already for $\frac{T}{T_c}\ge 4$ the equation of state is very close to that of 
a gas of free and massless particles ($P=\frac13\,\rho$): $\frac{T}{T_c}=4$ corresponds to 
$\kappa=\frac{P}{\rho}=0.3201$ and $\frac{T}{T_c}=6.2$ to $\kappa=\frac{P}{\rho}=0.3297$. 

\subsection{Value of $T_c$ for SU(2)$_{\tiny\mbox{CMB}}$\label{Tc}}

The strongest indication that $T_c$ of SU(2)$_{\tiny\mbox{CMB}}$, the Yang-Mills theory postulated 
to fundamentally describe photon physics, essentially coincides with the present temperature $T_0=2.725\,$K \cite{Cobe} 
of the CMB arises from observations of the so-called cosmic radio background, also dubbed {\sl unexplained extragalactic emission}, see 
\cite{Arcade2} and references therein or for a recent analysis \cite{JCAP2014}. In \cite{radioHofmann2009} the bump of spectral 
power, which exhibits a strong deviation from the quadratic dependence on frequency of the Planckian blackbody spectral radiance within the 
Rayleigh-Jeans regime, was interpreted as a re-shuffling effect of spectral power due to 
very low-frequency modes becoming evanescent: A Meissner induced, effective photon mass $m_\gamma$ induces a 
spectrum of standing waves at very low frequencies\footnote{At $T_c$ {\sl electric} (and unresolved) monopoles and antimonopoles 
\cite{SpatialWilson2009} start to 
condense \cite{HofmannBook2012}, indicating the onset of dynamical breaking of the so far unbroken U(1) gauge symmetry in 
the deconfining phase.}, assuming its maximum at zero. Modelling this in terms of a 
Gaussian shape, we extract the feeble value of present, effective photon mass of 
$m_\gamma\sim 100\,$MHz. Here the term {\sl effective} photon mass refers to the fact 
that the present Universe does not exhibit a homogeneous monopole condensate 
but rather is characterised by alternating spatial patches of deconfining and 
preconfining ground states which tunnel into one another. The quantity 
$m_\gamma$ thus represents an average over many of these patches, and there is not 
yet a (stable) third polarisation of the photon. Due to a critical increase of $m_\gamma$ when lowering $T$ away from $T_c$ 
such a small value for $m_\gamma$ implies that, practically, the present temperature $T_0=2.725\,$K of the CMB coincides 
with $T_c$. This determines the Yang-Mills scale $\Lambda$ of SU(2)$_{\tiny\mbox{CMB}}$ 
as $\Lambda=\frac{2\pi T_0}{\lambda_c}=\frac{2\times 2.725\pi}{13.87}\,\mbox{K}=1.234\,$K.

\subsection{Remark on polarisation tensor of thermal photon \label{Pi}}

In \cite{HofmannBook2012,HofmannNature,Hofmann2013} SU(2)$_{\tiny\mbox{CMB}}$ predictions, arising from a nontrivial 
photon polarisation tensor $\Pi_{\mu\nu}$, for thermal photon propagation (transverse part) and the induction of magnetic charge density waves 
(longitudinal part) at low temperatures and frequencies are given. Briefly, the transverse part of $\Pi_{\mu\nu}$ 
induces a screening-antiscreening modification of the low frequency part of the spectral, thermal energy density, 
in turn predicting the emergence of cosmologically local temperature depressions (redshift $z\sim 1$). This is a manifestation of dynamical 
statistical isotropy (and spatial homogeneity) breaking, exhibited through the temperature gradient 
seen by an observer located within a given depression. The longitudinal part repesents a cut off spectrum of 
longitudinally propagating, thermalised magnetic field modes in various branches 
\cite{Falquez2012}. To turn these into a magnetic field, coherent on cosmological length scales, requires a break of 
spatial homogeneity (local temperature gradient), as provided by 
the transverse part of $\Pi_{\mu\nu}$. This scenario, worth exploring in more detail, 
could yield a viable, unified description of the CMB large-angle anomalies (transverse part of $\Pi_{\mu\nu}$) 
\cite{WMAP,PlanckAno,LA} and the emergence of intergalactic magnetic fields (longitudinal part of $\Pi_{\mu\nu}$, coherence 
assisted by transverse part), both constituting late-time phenomena ($z\sim 1$). 
For what follows, small radiative corrections, such as described by $\Pi_{\mu\nu}$, are (and safely can be) neglected. 

\section{Low-$z$ and high-$z$ scaling of CMB temperature in SU(2)$_{\tiny\mbox{CMB}}$\label{tvsahighz}}

In this section we explore more basic cosmological consequences of 
SU(2)$_{\tiny\mbox{CMB}}$ than those related to the polarisation tensor $\Pi_{\mu\nu}$ 
discussed in Sec.\ref{Pi}. Namely, we ask what the implications are of Secs.\,\ref{Pandrho}, \ref{effe}, \ref{eossu2cmb}, 
and \ref{Tc} for (i) the dependence of CMB temperature $T$ on the cosmological 
scale factor, (ii) the cosmological evolution of SU(2)$_{\tiny\mbox{CMB}}$ energy density compared to that of a 
conventional photon gas, (iii) a resolution of the tension between the redshift value 
$z_{\tiny\mbox{re}}$ of instantaneous reionisation as extracted from the CMB TT angular power spectrum and 
Baryon Acoustic Oscillations (BAO) ($z_{\tiny\mbox{re}}=11.3\pm 1$) on one hand \cite{PlanckO,PlanckCP} 
and the detection/non-detection of the Gunn-Peterson trough in high-redshift quasar spectra ($z_{\tiny\mbox{re}}\sim 6$)
 on the other hand \cite{Quasar}, and (iv) the value of 
redshift $z_{\tiny\mbox{dec}}$ at CMB decoupling and the present cosmological concordance 
model at high redshift.     

\subsection{Energy conservation in an expanding Universe\label{eneconsSEC}}

We start by assuming SU(2)$_{\tiny\mbox{CMB}}$ to be a separately conserved 
cosmic fluid, stretched by the expansion of a Friedmann-Lema\^itre-Robertson-Walker Universe. The latter is characterised by the 
scale factor $a$ which we normalise to unity at present: $a(T_0)\equiv a_0=1$. In exclusively discussing the evolution of the 
CMB subsequent to decoupling, no specific cosmological model is required. In particular, no assumption on the Universe's spatial curvature needs to 
be made. Separate conservation of SU(2)$_{\tiny\mbox{CMB}}$ predicts interesting consequences for reionisation and 
CMB decoupling which are modified when this assumption 
is relaxed. This is done in Sec.\,\ref{neutrinomassproptoT} where we postulate the cosmic neutrino 
background to be conserved together with SU(2)$_{\tiny\mbox{CMB}}$ only. Interestingly, however, there are no 
essential modifications of these predictions under such an extension of SU(2)$_{\tiny\mbox{CMB}}$. 

Let $\rho$ and $P$ denote energy density and pressure of 
SU(2)$_{\tiny\mbox{CMB}}$, respectively. The equation of energy conservation reads 
\eqb
\label{enecons}
\frac{d\rho}{da}=-\frac{3}{a}(P+\rho)\,.
\eqe
To solve Eq.\,(\ref{enecons}) for $\rho(a)$, an equation of state $P=P(\rho)$ and a boundary condition $\rho^*=\rho(a^*)$ need 
to be prescribed. The former is obtained by solving $\rho=\rho(T)$ for $T=T(\rho)$ to be substituted 
into $P=P(T)$. The choice of initial condition is explained in Sec.\,\ref{T(a)}.    

\subsection{Temperature vs. scale factor and energy density of SU(2)$_{\tiny\mbox{CMB}}$\label{T(a)}}

\begin{figure}
\begin{center}
\leavevmode
\leavevmode
\vspace{4.9cm}
\includegraphics{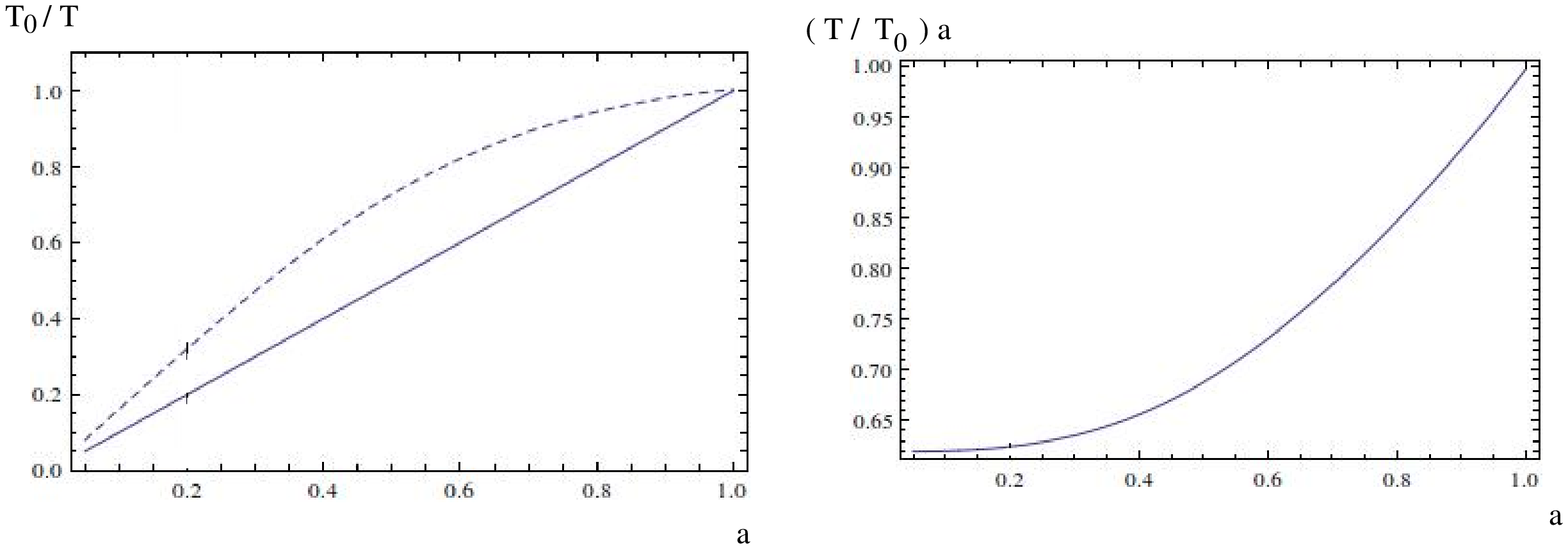}
\end{center}
\caption{\protect{\label{Fig-3}} SU(2)$_{\tiny\mbox{CMB}}$ induced violation of (conformal) U(1) scaling of 
CMB temperature $T$ with inverse scale factor $a^{-1}$. Left panel: $\frac{T_0}{T}$ as a 
function of scale factor $a$ for SU(2)$_{\tiny\mbox{CMB}}$ (dashed line) and for the  
conventional U(1) theory (solid line). Right panel: $\frac{T}{T_0}\times a$ as a function of 
scale factor $a$ for SU(2)$_{\tiny\mbox{CMB}}$. Note saturation of $\frac{T}{T_0}\times a$ to the value 
0.62 for $a<1/10$.}      
\end{figure}

We would like to derive an SU(2)$_{\tiny\mbox{CMB}}$ prediction for $(T_0/T)(a)$. To do this, 
the initial temperature $T^*$ (and hence energy density 
$\rho(T^*)$) is chosen such that, with a prescribed value $a^*<1$ and using $T=T(\rho(a))$,  
the evolution $\rho(a)$ generates the value 
$(T_0/T)=1$ at $a(T_0)=1$ (today). Prescribing $a^*=1/10$, one 
obtains $T^*=6.2\,T_0$, see left panel in Fig.\,\ref{Fig-3}. Judging from the results of 
Sec.\,\ref{eossu2cmb} and by saturation of $\frac{T}{T_0}\times a$ for $a<1/10$ (right panel of Fig.\,\ref{Fig-3}), 
it is safe to (conformally) scale $T$ with $a^{-1}$ for $a<1/10$. This is also expressed by the fact that the constant 
term 0.62 in fits of $\frac{T}{T_0}\times a$ to polynomials in $a$ is stable under variations of the fit interval, 
contained in $\frac{1}{20}\le a\le 1$, and the polynomial degree. 
Thus we have
\eqb
\label{Tscaling}
T=0.62\,a^{-1}\times T_0\,,\ \ \ \ (a\le \frac{1}{10})\,.
\eqe
Eq.\,(\ref{Tscaling}) and Fig.\,\ref{Fig-3} state that for $a<1$ the conformal scaling 
\eqb
\label{U(1)Tscaling}
T=a^{-1}\times T_0
\eqe
{\sl over-estimates} the {\sl actual} CMB temperature if it is SU(2)$_{\tiny\mbox{CMB}}$ that describes CMB photons correctly. 
Heuristically, this is a consequence of the fact that SU(2)$_{\tiny\mbox{CMB}}$ energy density $\rho$, 
chiefly residing in relativistic degress of freedom at high redshift, is not only reduced by spacetime growth,  
as in the conventional U(1) theory, but also by an investment into thermal ground-state structure and 
quasiparticle mass $m$. Recall that the former generates the latter by an increasingly efficient, adjoint Higgs mechanism 
as temperature drops towards $T_c$ (or $T_0$), see Fig.\,\ref{Fig-2}. 

Eq.\,(\ref{Tscaling}) implies the ratio $R_\rho$ of energy density $\rho$ in 
SU(2)$_{\tiny\mbox{CMB}}$ to energy density $\rho_\gamma=\frac{\pi^2}{15}\,T^4$ 
of a conventional photon gas to be 
\eqb
\label{Rrho}
R_\rho\equiv \frac{\rho}{\rho_\gamma}=0.591\ \ \ \ \ \ (a\le\frac{1}{10})\,.
\eqe
Thus, although at high temperatures ($a\le\frac{1}{10}$) SU(2)$_{\tiny\mbox{CMB}}$ possesses four times 
as many relativistic degrees of freedom as the conventional U(1) theory, 
its energy density $\rho$ is considerably smaller than $\rho_\gamma$. 
Based on Eq.\,(\ref{Tscaling}), we present in Sec.\,\ref{earlyre} observational evidence that, indeed, 
SU(2)$_{\tiny\mbox{CMB}}$ appears to yield a better description 
of the CMB than conventional U(1) theory.    

\subsection{The issue of early reionisation\label{earlyre}}

Due to nonlinear structure growth at late times hydrogen gas is compacted under gravitational pull, and stars come into being. 
Due to their radiation, the Universe's intergalactic medium suffers reionisation. A priori, there is no compelling 
reason why on cosmological time scales this process should {\sl not} be considered instantaneous: 
The large-scale {\sl distribution} of galaxies is homogeneous, and so, statistically speaking, each given and sufficiently large 
region of space experiences ionisation of its hydrogen like any other region of similar extent does. This should preclude sizable 
retardation effects. Observationally, cosmologically instantaneous reionisation is supported by the rapid transition 
in the $z$ dependence of quasar spectra. In \cite{Quasar} a moderate resolution Keck spectroscopy of 
quasars at $z=5.82, 5.99, 6.28$, discovered
by the Sloan Digital Sky Survey (SDSS), was performed. While the two objects with $z=5.82, 5.99$ do not 
exhibit the Gunn-Peterson trough, the object of highest redshift $z=6.28$ cleary does so, suggesting 
that reionisation indeed is a rapid transition occuring at $z_{\tiny\mbox{re}}\sim 6$. On the other hand, 
the latest value of a CMB based (WMAP and Planck) and Baryon Acoustic Oscillations (BAO) supported 
extraction of $z_{\tiny\mbox{re}}$ for instantaneous reionisation is $z_{\tiny\mbox{re}}\sim 11.3\pm 1.1$, 
see Table 10 (last column) of \cite{PlanckO} or Table 5 (last column) of \cite{PlanckCP}. Thus there is obvious tension between the 
results for $z_{\tiny\mbox{re}}$ from quasar spectra and those extracted from the CMB (plus BAO). 

Let us now address this discrepancy. Quasar light propagates with  
energy densities orders of magnitude higher than 
that of the CMB. Hypothetically equating such energy densities with $\rho$ of 
SU(2)$_{\tiny\mbox{CMB}}$, high fictitious temperatures $T$ would arise, $T\gg T_0$. 
But at such high temperatures the mean photon energy is just conformally redshifted as in the 
conventional U(1) theory. As a consequence, $z_{\tiny\mbox{re}}\sim 6$ of \cite{Quasar} 
should be trusted at face value as a {\sl physical} redshift for instantaneous reionisation: 
$z_{\tiny\mbox{re}}=6$  $\Rightarrow$ $a_{\tiny\mbox{re}}^{-1}=7$. 

We now test the validity of 
SU(2)$_{\tiny\mbox{CMB}}$ by appealing to the conventional assumption of 
conformal U(1) scaling as in Eq.\,(\ref{U(1)Tscaling}) to deduce the hypothetical value of 
$T_{\tiny\mbox{re}}$ associated with $z_{\tiny\mbox{re}}=6$. This assumption is underlying past and 
present CMB analysis, and in particular, the extraction of redshift for instantaneous reionisation in \cite{PlanckCP}. 
According to Eq.\,(\ref{U(1)Tscaling}) and accepting $z_{\tiny\mbox{re}}=6$ as physical,  
we conclude that $T_{\tiny\mbox{re}}=7\times T_0$. But if nature indeed realises 
SU(2)$_{\tiny\mbox{CMB}}$ then this value of $T_{\tiny\mbox{re}}$ would, 
according to Eq.\,(\ref{Tscaling}) translate into
\eqb
\label{cmbarec}
a_{\tiny\mbox{re}}=0.62\,\frac{T_0}{T_{\tiny\mbox{re}}}=0.0886\ \ \ \Rightarrow \ \ \ 
z_{\tiny\mbox{re}}=a_{\tiny\mbox{re}}^{-1}-1=10.29\,.
\eqe
Within errors the thus determined value of $z_{\tiny\mbox{re}}=10.29$ is consistent with
 $z_{\tiny\mbox{re}}\sim 11.3\pm 1.1$ obtained from combined (conventional) CMB and BAO analysis (Planck plus WMAP, high $l$ and BAO) 
\cite{PlanckCP}. Using the Planck data alone and invoking 
gravitational lensing, a somewhat lower value of $z_{\tiny\mbox{re}}$, subject to larger 
errors, was obtained by the Planck collaboration \cite{PlanckCP}: $z_{\tiny\mbox{re}}=10.8\pm^{3.1}_{2.5}$. Again, this is consistent 
with Eq.\,(\ref{cmbarec}). Reasoning in this way, the discrepancy in the values of $z_{\tiny\mbox{re}}$ is 
suggested to arise due to incorrect conformal U(1) scaling of CMB temperature when nature actually 
realises SU(2)$_{\tiny\mbox{CMB}}$. 

It is worth mentioning that the SU(2)$_{\tiny\mbox{CMB}}$ value of $T_{\tiny\mbox{re}}$ is 
$T_{\tiny\mbox{re}}=4.35\times T_0=11.85\,$K. In Sec.\,\ref{Masslessneutr} we will see that corrections to 
 Eq.\,(\ref{cmbarec}) due to (unconventional) neutrino physics are not severe.      

\subsection{Redshift at CMB decoupling\label{redec}}

The CMB decoupling temperature $T_{\tiny\mbox{dec}}$ of about $T_{\tiny\mbox{dec}}=3000\,$K (for our purposes it is 
sufficient to assume that recombination and CMB decoupling occur instantaneously and simultaneously) is 
unaffected\footnote{This value of $T_{\tiny\mbox{dec}}$ is a consequence of the ionisation energy of hydrogen, 
$E_{\tiny\mbox{ion}}=13.6\,$eV, and the Saha equation.} 
by SU(2)$_{\tiny\mbox{CMB}}$. According to Eq.\,(\ref{Tscaling}) $T_{\tiny\mbox{dec}}$ translates into a redshift 
$z_{\tiny\mbox{dec}}$ at decoupling of 
\eqb
\label{zdecpuresu2}
z_{\tiny\mbox{dec}}=\frac{1}{0.62}\,\frac{3000}{2.725}-1=1775\,. 
\eqe
This is substantially larger than the conventional value of $z_{\tiny\mbox{dec}}\sim 1100$ 
\cite{PlanckCP} and should have an impact on cosmological parameter values, notably matter 
density\footnote{Owing to SU(2)$_{\tiny\mbox{CMB}}$, matter density in the conventional 
concordance model is at CMB 
decoupling $\left(\frac{1775}{1100}\right)^3\sim 4.2$ times higher facing the same conventional photon pressure.} 
and the Hubble parameter, $H_0$. A detailed investigation of this important problem 
is well beyond the scope of the present work. 

\section{Massless neutrinos?\label{Masslessneutr}}

\subsection{Adaption of standard treatment of neutrino temperature to SU(2)$_{\tiny\mbox{CMB}}$\label{adapt}}

To start with, let us assume that neutrinos are massless and that there is no coupling between 
SU(2)$_{\tiny\mbox{CMB}}$ and the neutrino sector such that they represent separately conserved cosmic fluids. 
The standard argument of a conserved entropy 
density in the process of $e^+ e^-$ annihilation produces a ratio of neutrino temperature $T_\nu$ 
to CMB temperature $T$ of
\eqb\label{TneutTcmb}
\frac{T_\nu}{T}=\left(\frac{g_1}{g_0}\right)^{1/3}\,,
\eqe
where $g_0$ ($g_1$) denotes the number of relativistic degrees of freedom before (after)
$e^+ e^-$ annihilation. In the conventional theory one has: $g_0=2+\frac{7}{8}\,4$ and $g_1=2$ such that 
$\frac{T_v}{T}=\left(\frac{4}{11}\right)^{1/3}$. If we replace the conventional U(1) photon theory by 
SU(2)$_{\tiny\mbox{CMB}}$ then $g_0=8+\frac{7}{8}\,4$ and $g_1=8$ which yields
\eqb\label{TneutTcmbSU2}
\frac{T_\nu}{T}=\left(\frac{g_1}{g_0}\right)^{1/3}=\left(\frac{16}{23}\right)^{1/3}\,.
\eqe

\subsection{N$_{\tiny\mbox{eff}}\sim 3.36$ and separately conserved fluids of 
massless neutrinos\label{sepconsNeutr}}

As we have seen in Sec.\,\ref{T(a)}, it is safe to consider conventional, conformal scaling of $T$ versus $a^{-1}$ 
for $a\le 1/10$. Recall that $a=1/10$ corresponds to $T=6.2\,T_0$. Based on SU(2)$_{\tiny\mbox{CMB}}$ the effective number of neutrino 
flavours $N_{\tiny\mbox{eff}}$ at $T=T_0$, as judged by the conventional theory in terms of the actual number of massless neutrino 
flavours $N_\nu$, reads (total energy density in relativistic degrees of freedom minus energy density in photons 
divided by conventional energy density per massless neutrino flavour) 
\eqb
\label{neffnaive}
N_{\tiny\mbox{eff}}=\frac{\frac78\,N_\nu (0.62)^4 \left(\frac{16}{23}\right)^{4/3}}{\frac78 \left(\frac{4}{11}\right)^{4/3}}\,.
\eqe
Note that SU(2)$_{\tiny\mbox{CMB}}$ effects creep into the numerator of Eq.\,(\ref{neffnaive}) in terms of the 
factors $(0.62)^4$, related to the fact that $T_\nu$ of massless neutrinos always 
follows the conventional scaling law $T_\nu\propto a^{-1}$ while there are low-redshift violations 
thereof for $T$, and $\left(\frac{16}{23}\right)^{4/3}$, arising due to eight instead of two relativistic degrees of freedom in 
SU(2)$_{\tiny\mbox{CMB}}$ during $e^+ e^-$ annihilation (Sec.\,\ref{adapt}), deviating from unity and from $\left(\frac{4}{11}\right)^{4/3}$, respectively. 
For $N_\nu=3$ \cite{Z_0Lesgourges} we have $N_{\tiny\mbox{eff}}=1.053$ which is far off\footnote{If the (dimensionless) ground-state energy 
density of SU(2)$_{\tiny\mbox{CMB}}$, 
$T_0^{-4}\rho_{\tiny\mbox{gs}}(T_0)=32 \pi^4 \lambda_c^{-3}$, is added to the numerator of Eq.\,(\ref{neffnaive}) 
then one obtains $N_{\tiny\mbox{eff}}=6.2$. Clearly, this is also out of range. However, since (modulo small 
evanescence effects \cite{radioHofmann2009}) CMB photons decouple from their ground state at 
$T_0$ the ground-state part of SU(2)$_{\tiny\mbox{CMB}}$ at present should be viewed 
as a (tiny) contribution to dark energy rather than dark radiation. Therefore, we do not in the following
consider $\rho_{\tiny\mbox{gs}}(T_0)$ 
anymore when inferring $N_{\tiny\mbox{eff}}$ from SU(2)$_{\tiny\mbox{CMB}}$ and 
neutrino physics at present.} the observationally determined value of 
$N_{\tiny\mbox{eff}}\sim 3.36$ \cite{PlanckCP}. 

Interestingly, the value of $N_{\tiny\mbox{eff}}$ 
depends on the redshift at which it is determined. For example, one obtains for $a<1/10$
\eqb
\label{neffnaivea1/10}
N_{\tiny\mbox{eff}}=\frac{\frac78\,N_\nu\left(\frac{16}{23}\right)^{4/3}+3}{\frac78 \left(\frac{4}{11}\right)^{4/3}}\,.
\eqe
For $N_\nu=3$ we would have $N_{\tiny\mbox{eff}}=20.33$ instead of the value $N_{\tiny\mbox{eff}}=1.053$ extracted 
at present. From now on we associate $N_{\tiny\mbox{eff}}$ with its value today.  

\section{CMB thermalised neutrinos\label{neutrinomassproptoT}}

The results of Sec.\,\ref{sepconsNeutr} do suggest that 
SU(2)$_{\tiny\mbox{CMB}}$ and conventional neutrino physics are not 
compatible. Note that this extends to the case of neutrinos with fixed masses since the 
latter would reduce rather than enhance their contribution to 
the Universe's present energy density. Guided by results on how an SU(2) center-vortex responds 
to environmental conditions (putting forward an effective scale of resolution) 
\cite{Moosmann2008,MoosmannHofmann2008II} a bold suggestion on how to circumvent 
these difficulties is to assume that a given neutrino flavour is represented by a single center vortex loop of a 
respective SU(2) Yang-Mills theory and that this theory underwent its preconfing-confining phase 
transition well before CMB decoupling. Due to its extendedness and (after an electric-magnetically dual interpretation)  
its unit of {\sl electric} center flux such a center-vortex loop interacts with the 
CMB. As a consequence, neutrino temperature $T_\nu$ and CMB temperature $T$ would coincide: 
$T_\nu=T$ or $\left(\frac{16}{23}\right)^{1/3} \rightarrow 1$ in 
Eq.\,(\ref{TneutTcmb}). Also, no additional split of $T_\nu$ and $T$ at low $z$ due to a violation of 
conformal scaling in the photonic sector alone would occur.  

\subsection{The case of massless neutrinos}

Because it is technically simpler let us first assume that 
neutrinos, due to their interactions with the CMB, exhibit the same temperature, 
$T_\nu=T$, but that they remain massless. In this idealisation Eq.\,(\ref{neffnaive}) modifies as  
\eqb
\label{neffnaivebetter}
N_{\tiny\mbox{eff}}=\frac{\frac78\,N_\nu}{\frac78 \left(\frac{4}{11}\right)^{4/3}}\,.
\eqe
With $N_\nu=3$ one obtains $N_{\tiny\mbox{eff}}=11.56$ which is much too high. To reduce $N_{\tiny\mbox{eff}}$ down 
to its physical value $N_{\tiny\mbox{eff}}=3.36$ neutrinos need to acquire mass through interactions with the CMB.     

\subsection{Temperature dependent neutrino mass}

Due to an SU(2) center vortex loop possessing electromagnetic properties only, it exclusively 
interacts with the photonic part of SU(2)$_{\tiny\mbox{CMB}}$. Modulo radiative corrections, 
which are inessential for the present discusssion, the thermal photon gas 
in SU(2)$_{\tiny\mbox{CMB}}$ is characterised by the single scale $T$. Thus, 
the response of the neutrino sector in terms of neutrino mass emergence 
must be such that $m_\nu=\xi\,T$ where $\xi$ is a dimensionless 
constant of order unity. For cosmic neutrinos 
we assume $m_\nu$ to be universal, that is, flavour independent. We have given arguments in \cite{HofmannNature} on the viability of such a 
universal, temperature dependent neutrino mass in view of the overclosure bound of $\sim 15$\,eV, see 
\cite{LesPas2014}. Namely, for $0\le\xi\le 10$ this bound 
is evaded up to $z\sim 10^4$ which is well before CMB decoupling. On the other hand, lower 
bounds for the sum of neutrino masses of the order $10^{-1}$\,eV \cite{Valle2012}, posed by the 
two scenari of mass hierarchy (normal and inverted) to explain neutrino oscillations, are invalid for 
the here proposed temperature dependent mass of cosmic neutrinos because they refer to neutrino 
environments which are largely disparate from the CMB (neutrino generation and propagation in 
long baseline reactor, atmospheric, and solar neutrino experiments). 

\begin{figure}
\begin{center}
\leavevmode
\leavevmode
\vspace{4.9cm}
\includegraphics{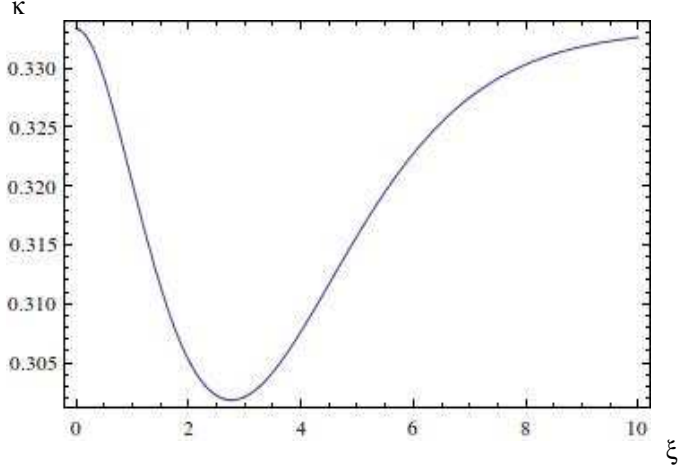}
\end{center}
\caption{\protect{\label{Fig-4}} The equation-of-state parameter $\kappa$ for the model of 
SU(2)$_{\tiny\mbox{CMB}}$ combined with $N_\nu=3$ temperature dependent massive neutrino flavours, 
see Eq.\,(\ref{neutrinoeandP}), as a function of $\xi$ for $T\gg T_0$. Note the smallness of the deviation from $\kappa=\frac13$, corresponding to 
the equation of state of a free gas of massless particles. The minimum of $\kappa$ (or maximal deviation from $\kappa=\frac13$) takes place 
at $\xi=2.77$. For comparison, the maximum of the blackbody spectral energy density occurs at $\frac{\omega}{T}=2.82$.}      
\end{figure}

With a universal, cosmic neutrino mass $m_\nu=\xi T$ the pressure $P_\nu$ and energy density $\rho_\nu$ 
is given as (2 spin orientations per flavour) 
\eab 
\label{neutrinoeandP}
P_\nu&=&N_\nu T^4\,\frac{1}{\pi^2}\,\int_0^\infty dx\,x^2\log(1 + \exp(-\sqrt{x^2 + \xi^2}))
\equiv N_\nu T^4\,\hat{P}_\nu(\xi)\,,\nonumber\\ 
\rho_\nu&=&N_\nu T^4\,\frac{1}{\pi^2}\,\int_0^\infty dx\,
\frac{x^2\sqrt{x^2 + \xi^2}}{1 + \exp\sqrt{x^2 + \xi^2}}\equiv N_\nu T^4\,\hat{\rho}_\nu(\xi)\,.
\eae
This model for the neutrino gas is not thermodynamically consistent by itself. 
(The CMB acts as a thermal background prescribing mass and inheriting its temperature to the neutrino gas.) 
Conservation of energy in an expanding Universe thus must be imposed onto the 
total energy density and pressure: $\rho\to\rho_{\tiny\mbox{tot}}=\rho+\rho_\nu$ and 
$P\to P_{\tiny\mbox{tot}}=P+P_\nu$ in Eq.\,(\ref{enecons}). The model of Eqs.\,(\ref{neutrinoeandP}) 
exhibits energy density and pressure which both are proportional to $T^4$ like the CMB does 
(the photonic part of SU(2)$_{\tiny\mbox{CMB}}$ or the entire 
SU(2)$_{\tiny\mbox{CMB}}$ for $T\gg T_0$). That is, for $T\gg T_0$, 
the ratio $\frac{P_{\tiny\mbox{tot}}}{\rho_{\tiny\mbox{tot}}}\equiv\kappa$ is independent of $T$. It reads
\eqb
\label{eos_tot}
\kappa=\frac{\frac13+\frac{15}{4\pi^2}N_\nu \hat{P}_\nu(\xi)}{1+\frac{15}{4\pi^2}N_\nu \hat{\rho}_\nu(\xi)}\,, \ \ \ (T\gg T_0)\,.
\eqe
Fig.\,\ref{Fig-4} depicts $\kappa$ as a function of $\xi$ for $T\gg T_0$ and $N_\nu=3$. 
Because of the small deviation $\epsilon\equiv\frac13-\kappa$, reaching its maximal value $\epsilon_{\tiny\mbox{max}}=0.0315$ at 
$\xi=2.77$, the high-temperature scaling of this combination of SU(2)$_{\tiny\mbox{CMB}}$ 
with $N_\nu=3$ temperature dependent massive neutrino flavours exhibits only weak deviations from conformal 
scaling,  
\eqb\label{devconscal}
\frac{T}{T_p}=a^{-1+\frac34 \epsilon}\,,\ \ \ \ \ (T\ge T_p;\, T_p\gg T_0;\, a(T_p)=1)\,,
\eqe
where $T_p$ denotes a pivotal temperature within the high-temperature regime. 

\begin{figure}
\begin{center}
\leavevmode
\leavevmode
\vspace{4.9cm}
\includegraphics{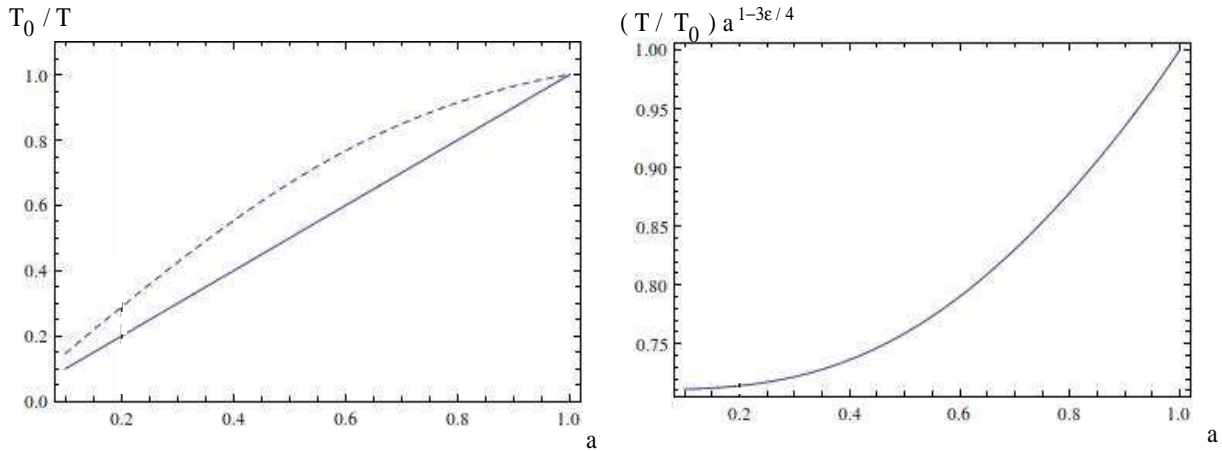}
\end{center}
\caption{\protect{\label{Fig-5}} Violation of conformal scaling as induced by 
SU(2)$_{\tiny\mbox{CMB}}$ combined with $N_\nu=3$ temperature dependent massive neutrino flavours. 
Left panel: $\frac{T_0}{T}$ as a 
function of scale factor $a$ for this model (dashed line) and for a 
conventional U(1) photon theory (solid line). Right panel: $\frac{T}{T_0}\times a^{1-\frac34\epsilon}$, 
compare with Eq.\,(\ref{devconscal}), as a function of 
scale factor $a$ for this model. Saturation to $\frac{T}{T_0}\times a^{1-\frac34\epsilon}=0.71$ occurs for 
$a<1/10$ which is consistent with $\frac{T}{T_0}=6.8$ at $a=1/10$ and $\epsilon=0.0263$.}      
\end{figure}

Let us now determine the value of $\xi$ such that todays's value of $N_{\tiny\mbox{eff}}(\xi)$, defined as
\eqb
\label{neffnaivebetterneutrinomass}
N_{\tiny\mbox{eff}}(\xi)=\frac{N_\nu\hat{\rho}_\nu(\xi)}{\frac78\frac{\pi^2}{15}\left(\frac{4}{11}\right)^{4/3}}\,,
\eqe
matches $N_{\tiny\mbox{eff}}=3.36$, the value observationally determined in \cite{PlanckCP}. We find
\eqb
\label{planckvalueofxi}
\xi=3.973\,,
\eqe
which is to the right of the minimum in Fig.\,\ref {Fig-4}. This corresponds to $\epsilon=0.0263$. 
In Fig.\,\ref {Fig-5} we show in analogy to Fig.\,\ref{Fig-3} how scaling violation occurs in SU(2)$_{\tiny\mbox{CMB}}$ 
combined with $N_\nu=3$ temperature 
dependent massive neutrino flavours ($\xi=3.973$). In contrast to the case of 
pure SU(2)$_{\tiny\mbox{CMB}}$ there is a violation of conformal U(1) scaling 
also for $T\gg T_0$. This is expressed by Eq.\,(\ref{devconscal}). 
. 

Employing Eq.\,(\ref{devconscal}) with $T_p=6.8\,T_0$, we now estimate, in analogy to Sec.\,\ref{redec}, 
the redshift $z_{\tiny\mbox{dec}}$ at CMB decoupling as
\eqb
\label{zdecsu2neutr}
z_{\tiny\mbox{dec}}=10\,\left(\frac{3000}{6.8\times 2.725}\right)^{\frac{1}{1-\frac34\epsilon}}-1=1793.5\,. 
\eqe
Interestingly, this value of $z_{\tiny\mbox{dec}}$ differs from $z_{\tiny\mbox{dec}}=1775$, obtained for the case of 
pure SU(2)$_{\tiny\mbox{CMB}}$ (see Eq.\,(\ref{zdecpuresu2})), by only 1\%.  

What about the redshift $z_{\tiny\mbox{re}}$ for instantaneous reionisation? The same reasoning as in 
Sec.\,\ref{earlyre} but now subject to the modified scaling of Eq.\,(\ref{devconscal}) yields the following expression
\eqb
\label{cmbarecneutrino}
a_{\tiny\mbox{re}}=\frac{1}{10}\,\left(\frac{6.8\,T_0}{T_{\tiny\mbox{re}}}\right)^{\frac{1}{1-\frac34\epsilon}} =0.0970862\ \ \ \Rightarrow \ \ \ 
z_{\tiny\mbox{re}}=a_{\tiny\mbox{re}}^{-1}-1=9.3\,,
\eqe
where $T_{\tiny\mbox{re}}=7\times T_0$. This value of $z_{\tiny\mbox{re}}$ remains compatible with the one 
extracted by the Planck collaboration using the CMB and gravitational lensing only, 
$z_{\tiny\mbox{re}}=10.8\pm^{3.1}_{2.5}$ \cite{PlanckCP}, it 
differs from $z_{\tiny\mbox{re}}=10.29$, obtained for the case of 
pure SU(2)$_{\tiny\mbox{CMB}}$ (see Eq.\,(\ref{cmbarec})), by $-$9.6\%, and it is at slight tension with the 
value $z_{\tiny\mbox{re}}\sim 11.3\pm 1.1$ obtained from combined (conventional) CMB and BAO analysis.

\section{Summary and Conclusions\label{Sum}}

In this work we have put into perspective consequences of the postulate that a pure SU(2) Yang-Mills 
theory, posessing a critical temperature $T_c$ for the deconfining-preconfining phase transition 
which coincides with the present temperature $T_0=2.725\,$K of the CMB \cite{HofmannBook2012}, describes the 
cosmological evolution of the CMB. The here addressed physics, which only appeals 
to the free quasiparticle level in describing thermodynamical quantities in 
the deconfining phase of SU(2)$_{\tiny\mbox{CMB}}$, is much more basic 
than SU(2)$_{\tiny\mbox{CMB}}$'s imprint on large-angle anisotropies due to particular radiative corrections \cite{HofmannNature}. 
We have also addressed how the SU(2)$_{\tiny\mbox{CMB}}$ scenario could affect 
cosmic neutrinos. An apparently viable scenario, where, by interacting with the CMB, cosmic neutrinos 
stream at CMB temperature $T$ and acquire 
a mass $\propto T$, does not qualitatively change the 
prediction for $z_{\tiny\mbox{dec}}$ (redshift at which CMB decouples) and for the redshift value 
$z_{\tiny\mbox{re}}$ of instantaneous reionisation, the latter simulated under a conventional, conformal scaling assumption {\sl and} 
SU(2)$_{\tiny\mbox{CMB}}$. This is done by conventionally relating the value 
$z_{\tiny\mbox{re}}\sim 6$, extracted from quasar spectra, to the value of $T$, and by 
subsequently computing from this the associated SU(2)$_{\tiny\mbox{CMB}}$ redshift. The latter happens to be 
compatible with the value extracted by the Planck collaboration using the CMB and 
gravitational lensing only.       

SU(2)$_{\tiny\mbox{CMB}}$ predicts a violation of the simple $T\propto a^{-1}$ scaling at 
low redshift which moves $z_{\tiny\mbox{dec}}$ from $z_{\tiny\mbox{dec}}\sim 1100$ to 
$z_{\tiny\mbox{dec}}\sim 1800$. To keep, at CMB decoupling, the ratio of thermal photon energy density  
to matter density unchanged, matter density would have 
to be rescaled by a factor $\left(\frac{1100}{1800}\right)^{3}\sim 0.23$ yielding a new $\Omega_m$ of 
about 7\,\%. This, however, is close to the present baryonic density ($\sim 5.5\,\%$). 
At high redshifts the validity of the 
$\Lambda$CDM concordance model is thus questioned. In a cosmological model void of dark matter, 
the strong $\Omega_m$ component seen in cosmologically 
local signatures (luminosity distance - redshift curves, large-scale structure surveys)  
could be an indication of the onset of coherent oscillations of a homogeneous 
dark-energy field at late times \cite{Planckscaleaxion}, and conventional 
dark-matter halos, thought to be responsible for the flattening of galaxy rotation curves, 
could be mimicked by (topologically stabilised) solitonic configurations of such a field.      

To rule out or strengthen the here proposed scenario for CMB and cosmic neutrino physics, 
dedicated simulations of the implied cosmological model are required resting on 
data from both local cosmology and the CMB angular correlation functions.

\section*{Acknowledgments}
We would like to acknowledge useful conversations with Julian Moosmann. We also thank 
the organisers of {\sl Cosmology and Fundamental Physics with Planck} at CERN (June 2013) for an insightful event. 
This paper is dedicated to the memory of {\sl Pierre van Baal} and {\sl Dmitri Diakonov} who both were 
exceptionally gifted and committed theoretical physicists, contributing important pieces to the development of 
Quantum Yang-Mills theory.

%
%
\end{document}